\documentclass[showpacs,preprintnumbers,amsmath,amssymb,nofootinbib]{revtex4-2}
\usepackage{graphicx}
\usepackage{color, soul}
\usepackage{amsmath}
\usepackage{amsfonts}
\input epsf
\usepackage[pdftex,
            pdfauthor={A. V. Yurov, V. A. Yurov},
            pdftitle={On the Question of the B\"acklund Transformations and Jordan Generalizations of the Second Painlev\'e Equation},
            pdfsubject={},
            pdfkeywords={Painlev\'e equations; B\"acklund transformations; Schlesinger transformations; JS-systems; JP-systems},
            pdfproducer={Latex with hyperref},
            pdfcreator={pdflatex}]{hyperref}
\hypersetup{colorlinks=true, citecolor=blue, linkcolor=red}
\usepackage{enumerate}
\usepackage{scrextend}
\addtokomafont{labelinglabel}{\sffamily}

\newcommand{\beq}{\begin{equation}}
\newcommand{\beqn}{\begin{equation*}}
\newcommand{\enq}{\end{equation}}
\newcommand{\enqn}{\end{equation*}}

\newcommand{\R}{{\mathbb R}}
\newcommand{\N}{{\mathbb N}}

\newcommand{\Ai}{\text{\rm Ai}}
\newcommand{\Bi}{\text{\rm Bi}}
\newtheorem{Proposition}{Proposition}
\newtheorem{Remark}{\textsc{Remark}}
\newtheorem{Example}{\textsc{Example}}

\newtheorem{thm}{Theorem}
\newtheorem{property}[thm]{Property}

\begin{document}
\title{On the Question of the B\"acklund Transformations and Jordan Generalizations of the Second Painlev\'e Equation}
\author{A.V. Yurov}%
\email{AIUrov@kantiana.ru}
\author{V.A. Yurov}%
\email{vayt37@gmail.com}
\affiliation{I. Kant Baltic Federal University, Department of Physics, Mathematics and IT, Al. Nevsky str. 14, Kaliningrad
236041, Russia}

\begin{abstract}
We demonstrate the way to derive the second Painlev\'e equation $P_2$ and its B\"acklund transformations from the deformations of the Nonlinear Schr\"odinger equation (NLS), all the while preserving the strict invariance with respect to the Schlesinger transformations. The proposed algorithm allows for a construction of Jordan algebra-based completely integrable multiple-field generalizations of $P_2$ while also producing the corresponding B\"acklund transformations. We suggest calling such models the JP-systems. For example, a Jordan algebra $J_{_{{\rm Mat}(N,N)}}$ with the Jordan product in the form of a semi-anticommutator is shown to generate an integrable matrix generalization of $P_2$, whereas the $V_{_N}$ algebra produces a different JP-system that serves as a generalization of the Sokolov's form of a vectorial NLS.
\end{abstract}

\maketitle

\section{Introduction}\label{Sec:intro}
\allowdisplaybreaks
The triumphant emergence of the Painlev\'e equations dates back to the very end of the nineteenth century, when Emile Picard posed the following question \cite{Picard}: what kind of second order O.D.E.s contain no movable singularities except for poles? By 1910, Paul Panilev\'e and his student Bertrand Gambier \cite{Painleve00,P02,G10} proved that there are 50 types of second order O.D.E.s with such a property. Six of them were shown to be irreducible to either elementary or classical special functions. These interesting solutions have been dubbed the Painlev\'e transcendents, and the corresponding six equations, the Painlev\'e equations I--VI. The Painlev\'e equations have been extensively studied as isomonodromic deformations of linear systems \cite{F05,F07,Sl12,GG12,GG17}, and to this day, they remain one of the most important ingredients of the integrable systems theory and the one most shrouded in mystery. The equations arise in the problems that involve self-similar solutions of integrable hierarchies \cite{ARS1,ARS2}; they helped establish a (to this day not sufficiently understood) relationship between the integrability and the problems of preservation of O.D.E.s' monodromy \cite{Jimbo,FlNe}; and they positively proliferate when one studies the dressing chains of discrete \mbox{symmetries \cite{VSh}}. Additionally, besides the multitude of articles that are dedicated to the equations themselves, there are also many works on the various generalizations of said equations, running the gamut from discrete to multiple-fields matrix models.

However, the importance of the Painlev\'e equations is not limited to the field of mathematics, since many of them arise in a number of interesting physical problems. For example, the solutions of Painlev\'e III emerge in the studies of the spin--spin correlation function in the 2D Ising model \cite{Wu76} and occur in the scaling functions for two-dimensional polymers \cite{Fen92,Z94}; Painlev\'e IV is a feature in generalizations of odd superpotentials when one studies the exact nonsingular cosmological solutions on a 3D brain 
that interacts with five-dimensional gravity and the bulk scalar field \cite{TemaLera}; Painlev\'e V is necessary in the descriptions of a density matrix of an impenetrable Bose gas \cite{Jim80}; and the conformal field theory was even shown to be intimately connected to Painlev\'e VI \cite{Gil12}, as well as to V and III \cite{Gil13} (see also \cite{Stchyot}).


In this article, we will concentrate on Painlev\'e II, which for brevity we will henceforth call $P_2$:
\begin{equation}
w''(x,\alpha)=2w^3(x,\alpha)+xw(x,\alpha)+\alpha.
\label{P_2}
\end{equation}

It, too, has a plethora of interesting applications, appearing, for instance, in the long time asymptotics for the Kardar--Parisi--Zhang  equation \cite{Amir10,Sas10} and in the descriptions of the one-dimensional asymmetric simple exclusion process (ASEP) on the integer lattice \cite{Tracy2010}.  In addition, the $P_2$ equation arises in various physical problems: for example, during the capture into resonance of two weakly connected nonlinear oscillators \cite{Kalyam} and in the framework of the electrostatic probe theory \cite{Kasha}. Thanks in part to all these applications, ever since its inception in \cite{Painleve00}, $P_2$ has been in a spotlight of many research projects, including ones that concerned themselves with the task of constructing the exact solutions for \eqref{P_2}. It is known, in particular, that (\ref{P_2}) has rational solutions as long as the parameter $\alpha$ remains an integer \cite{Airault,Luka}. It is also known that each of these solutions is represented by a logarithmic derivative of a pair of certain polynomials, known in the literature as the Yablonskii--Vorob'ev polynomials \cite{Yabl,Vorob,Clark1,Demina}. In \cite{Clarc2}, these polynomials were shown to crop up in general solutions of soliton P.D.E.s, from the Korteweg–De Vries (KdV) equation and NLS to Kadomtsev--Petviashvili (KP) equations (see also \cite{Demina}), while in \cite{Luka}, it has been demonstrated that {\em any} rational solution of $P_2$ can be constructed via the B\"acklund transformations (\ref{3.4})--(\ref{3.6}) (we will proceed to explicitly derive these transformations later in the article) from the solution to the homogeneous version of \eqref{P_2}, i.e., the one with  $\alpha=0$.

The article \cite{DefNLS} has introduced a powerful new way to derive the Painlev\'e equations from integrable evolution equations. Its gist was to utilize the deformations of NLS. First, one introduces an auxiliary linear evolution equation and one auxiliary spectral problem, in which the potential depends on the spectral parameter as a polynomial of either the first or second order. While the former polynomial leads to KdV, the latter one produces NLS, which is natural, since a quadratic dependence upon the spectral parameter is but a linear Zakharov--Shabat problem, albeit written in some special gauge \cite{Sh-3}. In the next step, one performs a deformation of both the spectral problem and of the compatibility equation. Finally, one looks for the travelling wave solutions of the resulting equations, only to end up with the sought after Painlev\'e equation. It is a very attractive method, but it is also admittedly somewhat overwrought. And as we shall see below, it is quite possible to derive $P_2$ from the (trivially) deformed NLS system in a more direct and natural way; it appears that the key to this lies not in the auxiliary spectral problem, but in the famous symmetries of NLS known as Schlesinger transformations (a term we use following the seminal work \cite{Newell}). Furthermore, it is these very transformations that end up serving as the B\"acklund transformations for $P_2$. Even better, this approach is versatile enough to be easily generalizable to multiple-fields models. To be more precise, recall that all multiple-fields integrable NLS generalizations can be embedded into one general formalism, associated with a unital Jordan algebra \cite{Svin}. For example, if one starts with a Jordan algebra $J_{_{{\rm Mat}(N,N)}}$ with the Jordan product in the form of an anticommutator divided by two, it will lead to a system of matrix-valued NLS equations. Such systems are called the JS-systems (where JS stands for Jordan--Shr\"odinger), and since we shall soon demonstrate that our approach naturally leads to a matrix-valued analogue of $P_2$ (and to its B\"acklund transformations), we think it natural to introduce the term {\bf JP-systems} for such equations (JP here stands for Jordan--Painlev\'e). While the matrix-valued Painlev\'es themselves have been a subject of many research papers, our method clearly delineates them as but the first and simplest exhibit in a larger JP-systems menagerie. Other types of Jordan algebras will instead produce their own distinct JP-systems.

Before we continue, let us briefly discuss the structure of this paper. Section \ref{Sec:Schelsinger} is dedicated to a discussion of relevant symmetries, known as the Schlesinger transformations for NLS, and to their relationship with the Toda chain equations. We delve deeper in Section \ref{Sec:deformations}, where we study the NLS deformations, prove the invariance of the trivial deformations with respect to the Schlesinger transformations, and then derive the system of homogeneous split $P_2$ equations. This system is equivalent to a single fourth order O.D.E., but its order can be lowered, and it is exactly what we will do in Section \ref{Sec:matrices}. There, we derive the complete $P_2$ equation along with the corresponding B\"acklund transformations and show a simple way to generalize the results for the matrix-valued analogue of $P_2$. While we are at it, we also demonstrate how a number of this equation's properties, originally proven only via rather difficult calculations, turn out to be essentially self-evident when viewed through the lens of our approach. Similarly, deriving the B\"acklund transformations for the matrix-valued $P_2$ is very easy and is accomplished at the end of that Section. After that, we move on to Section \ref{Sec:vectors}, where we describe a new JP-system which serves as a generalization of a vectorial Sokolov model.
In Section \ref{Sec:P4}, we take a small step out of the confines of the discussion focused on Painlev\'e II and very briefly describe the first few steps in a possible application of our method to the task of Jordan generalizations of Painlev\'e IV ($P_4$).
Finally, we conclude this paper in Section \ref{Sec:conclusion} by briefly discussing some of the most straightforward and potentially interesting steps to further the subsequent development of this rich topic.

\section{Schlesinger Transformation} \label{Sec:Schelsinger}

Consider a split Nonlinear Schr\"odinger equation (NLS):
\begin{equation}
\begin{split}
u_t &= u_{xx} + 2u^2 v, \\
-v_t &= v_{xx} + 2v^2 u,
\end{split}
\label{1.1}
\end{equation}
where $u=u(x,t)$, $v=v(x,t)$.
A Schlesinger transformation (ST) for \eqref{1.1} has the form
\begin{equation}
u\to u_1=\frac{1}{v},\qquad v\to v_1=v\left(uv+(\log v)_{xx}\right),
\label{1.2+}
\end{equation}
and
\begin{equation}
 u\to u_{-1}=u\left(uv+(\log u)_{xx}\right),\qquad v\to v_{-1}=\frac{1}{u}.
\label{1.2-}
\end{equation}

It is easy to see that
\begin{equation}
 \left(u_1\right)_{-1}=\left(u_{-1}\right)_1=u,\qquad \left(v_1\right)_{-1}=\left(v_{-1}\right)_1=v,
\label{1.3}
\end{equation}

Hence (\ref{1.2+}) and (\ref{1.2-}) are explicitly invertible B\"acklund auto-transformations for \eqref{1.1}.

\begin{Remark} \label{Rem:AKNS}
The article \cite{AKNS} has introduced the concept of the so-called Ablowitz-Kaup-Newell-Segur (AKNS) hierarchy of nonlinear integrable P.D.E.s. The hierarchy starts with a (split) NLS, then goes to the coupled modified Korteweg--de Vries (cmKdV) system, and the third entry is the Lakshmanan--Porsezian--Daniel (LPD) equations.~(For a detailed description of the AKNS hierarchy, see \cite{MS}.) It is possible to prove by induction that all nonlinear equations belonging to the AKNS hierarchy are invariant with respect to the Schlesinger transformations (\ref{1.2+}) and (\ref{1.2-}). However, the proof is quite cumbersome and lies out of scope of this work and, as such, will be the subject of a follow-up article.
\end{Remark}

The STs are intimately related to the Toda chain equations. If we define
$$
s_n=\log u_n,\qquad n=-1,0,1,\qquad u_0=u.
$$
then the transformation
\beqn
u\to u_{-1}=u\left(uv+(\log u)_{xx}\right),
\enqn
coupled with the fact that
\beqn
v=u_1^{-1}={\rm e}^{-s_1}
\enqn
leads to a welcome surprise in the shape of Toda lattice:
\begin{equation}
s_0''=\exp\left(s_{-1} - s_0\right)-\exp\left(s_{0}-s_{1}\right),
\label{1.4}
\end{equation}
where the prime denotes the derivative with respect to the variable $x$. It is safe to say that the relationship between Equations (\ref{1.2+}) and (\ref{1.2-}), on the one hand, and the Toda chain (\ref{1.4}), on the other hand, is truly one of most peculiar properties of ST.

There are many known ways to generalize the Schlesinger's approach. For example, one might jump from the $(1+1)$  NLS (\ref{1.1}) to the $(1+2)$ Davey--Stewartson (DS) \mbox{equations \cite{Tema1}}. Alternatively, one might instead utilize the Jordan generalizations of NLS, introduced by Svinolupov and Yamilov \cite{Svin}; this approach is known to produce a number of non-trivial structures known as JS-systems and JT-systems. However, in this article we will concentrate on the third avenue of research that couples ST with $P_2$. In particular, we will demonstrate that the famous B\"acklund transformations for $P_2$ are {\bf exactly} the Schlesinger transformations (\ref{1.2+}) and (\ref{1.2-}). We will perform it by starting out with the NLS deformations.

\section{The NLS Deformations}\label{Sec:deformations}
The NLS deformation can be written in the following form \cite{DefNLS}:
\begin{equation}
\begin{split}
u_t =& c_1\left[x\left(u_{xx}+2u^2v\right)+2u_x+2uD_x^{-1}(uv)\right]+c_2\left(u_{xx}+2u^2v\right)+\\
&+ c_3(xu)_x+c_4xu+c_5u_x,
\label{2.1}
\end{split}
\end{equation}
\begin{equation}
\begin{split}
v_t =& -c_1\left[x\left(v_{xx}+2v^2u\right)+2v_x+2vD_x^{-1}(uv)\right]-c_2\left(v_{xx}+2v^2u\right)+\\
 &+ c_3(xv)_x-c_4xv+c_5v_x,
\label{2.2}
\end{split}
\end{equation}
where $D_x^{-1}$ denotes an indefinite integral with respect to the variable $x$. As has been noticed by Shabat and Yamilov, these equations are related to four of the Painlev\'e equations: $P_2$, $P_3$, $P_4$, and $P_5$.

Let $c_1=0$, $c_2\ne 0$. Then the deformations (\ref{2.1}) and (\ref{2.2}) are trivial and reducible to the split NLS (via the shift- and Galilean point transformations). In this case, (\ref{2.1}) and (\ref{2.2}) are invariant with respect to ST, coupled with the point transformations. The latter can be taken care of by setting $c_3=0$. This yields the following:

\begin{Proposition} \label{Proposition} Equations (\ref{2.1}) and (\ref{2.2}) with $c_1=c_3=0$ are invariant with respect to ST (\ref{1.2+}), (\ref{1.2-}), regardless of coefficients $c_2$, $c_4$, and $c_5$.
\end{Proposition}

Consider a travelling wave solution of (\ref{2.1}) and (\ref{2.2}):
$$
u=u(x+c_5t),\qquad v=v(x+c_5t),
$$
which abides by the following conditions:
$$
c_4=-c_2,\qquad c_5=0,
$$
in accordance with Proposition \ref{Proposition} (keep in mind that $c_1=c_3=0$). After the point transformation $u\to -u$ and assuming that $c_2\ne 0$, systems (\ref{2.1}) and (\ref{2.2}) reduce to
\begin{equation}
u''=\left(2uv+x\right)u,\qquad v''= \left(2uv+x\right)v,
\label{2.3}
\end{equation}
which can be called a {\em split homogeneous $P_2$ equation}, because when $u=v=w(x,0)\equiv w$, it further reduces to a special case of \eqref{P_2} with $\alpha=0$.

Of course, the new {\em split $P_2$}
\beqn
u''=\left(2uv+x\right)u,\qquad v''= \left(2uv+x\right)v,
\enqn
is invariant with respect to ST (\ref{1.2+}) and (\ref{1.2-}), which, for this case, take the form
\begin{equation}
u\to u_1=\frac{1}{v},\qquad v\to v_1=\frac{v'^2}{v}-(uv+x)v,
\label{2.5}
\end{equation}
\begin{equation}
u\to u_{-1}=\frac{u'^2}{u}-(uv+x)u,\qquad v\to v_{-1}=\frac{1}{u},
\label{2.6}
\end{equation}
and, naturally, $\log u_n$, $\log v_n$ still satisfy the Toda chain (\ref{1.4}).

At first glance, the results we have gained are less than spectacular. First of all, it is easy to see that the new STs (\ref{2.5}) and (\ref{2.6}) {\em do not} preserve the reduction $u=v$, thus creating the impression that the equation $P_2$ \eqref{P_2} might, after all, be unrelated to both (\ref{2.5}) and (\ref{2.6}) and the corresponding Toda chain. Secondly, the r.h.s. of (\ref{2.3}) bears no constant terms and thus might at best be reducible to a homogeneous $P_2$ with $\alpha=0$. All of this is exacerbated by the fact that, in general, the equation we end up with will be of a {\bf fourth} order:
\begin{equation}
{\bf L}[u]\equiv u^2u''''-4uu'u'''-3u(u'')^2+2u''(3u'^2+xu^2)+2uu'(u-xu')=0.
\label{2.7}
\end{equation}

However, as we shall see below, the order of (\ref{2.7}) can be reduced. Before we do that, though, we shall point out a number of very interesting properties it possess.
\begin{property} \label{Prop1}
Equation (\ref{2.7}) is invariant with respect to transformation $u\to \beta u$, with $\beta={\rm
const}$:
$$
{\bf L}[\beta u]=\beta^3 {\bf L}[u]=0.
$$
\end{property}

\begin{property} \label{Prop2}
Equation (\ref{2.7}) is invariant with respect to three discrete transformations, generated by the nonlinear operators ${\bf P_0}$,  ${\bf P_1}$, ${\bf P_2}$:
\begin{equation}
{\bf P_0} u=\frac{1}{u}, \label{p0}
\end{equation}
\begin{equation}
{\bf P_1} u=\frac{u''-xu}{2u^2}, \label{p1}
\end{equation}
\begin{equation}
{\bf P_2} u=\frac{2u}{2u'^2-uu''-xu^2}, \label{p2}
\end{equation}
where $u$ is a solution of 4th order Equation (\ref{2.7}).
\end{property}

\begin{property} \label{Prop3}
The operators ${\bf P}_0$, ${\bf P}_1$, and ${\bf P}_2$ have the following properties:
\begin{equation}
{\bf P_0}^2={\bf P_1}^2={\bf P_2}^2=E, \label{pr-1}
\end{equation}
where $Eu=u$ and
\begin{equation}
\begin{array}{l}
{\bf P_1}{\bf P_0}={\bf P_0}{\bf P_2},\\
{\bf P_0}{\bf P_1}={\bf P_2}{\bf P_0}.
\end{array}
 \label{pr-2}
\end{equation}
\end{property}

\begin{Remark}\label{Rem:P0_P1_P2} The interweaving relations (\ref{pr-2}) are nontrivial because
$$
\left[{\bf P_0},{\bf P_{1,2}}\right]\ne 0.
$$

For example,
$$
\left[{\bf P_1},{\bf
P_{0}}\right]=\frac{2u'^2(u''-xu)-u({u''}^2+4u^2-x^2u^2)}{2u(u''-xu)}.
$$
Using these properties one can construct an infinite set of exact solutions of (\ref{2.7}), building upon some simple seminal solution.
\end{Remark}

\begin{Example}
 Let us begin by adopting a trivial solution $u_0=1$ of (\ref{2.7}). We obtain an infinite set of progressively more complex solutions from it. Let us list some of them:
\begingroup\makeatletter\def\f@size{8.5}\check@mathfonts
\def\maketag@@@#1{\hbox{\m@th\normalsize \normalfont#1}}%
\beq \label{u1-u6}
\begin{split}
u_1 &={\bf P_1}u_0=-\frac{x}{2}, \\
u_2 &={\bf P_2}{\bf P_1}u_0=\frac{4x}{x^3-2}, \\
u_3 &={\bf P_1}{\bf P_2}{\bf P_1}u_0=-\frac{x^6-10x^3-20}{8(x^3-2)},\\
u_4 &=\left({\bf P_2}{\bf P_1}\right)^2u_0=\frac{16(x^6-10x^3-20)}{x(x^9-30x^6-1400)}, \\
u_5 &={\bf P_1}u_4=-\frac{x^{15}-70x^{12}+700x^9-9800x^6-196000x^3+196000}{32x(x^9-30x^6-1400)}, \\
u_6 &=\left({\bf P_2}{\bf P_1}\right)^3u_0=\\
&\frac{64(x^{15}-70x^{12}+700x^9-9800x^6-196000x^3+196000)}
{x^{21}-140x^{18}+4620x^{15}-78400x^{12}-1078000x^9-45276000x^6+301840000x^3+301840000},
\end{split}
\enq
\endgroup
\newline
and so on.
\end{Example}

\begin{Remark} \label{Prop4}
If $\lambda$ is constant, then
\begin{equation}
\begin{array}{l}
\displaystyle{
{\bf P_0}\left(\lambda u\right)=\frac{1}{\lambda}{\bf P_0}u,}\\
\\
\displaystyle{
{\bf P_1}\left(\lambda u\right)=\frac{1}{\lambda}{\bf P_1}u,}\\
\\
\displaystyle{ {\bf P_2}\left(\lambda
u\right)=\frac{1}{\lambda}{\bf P_2}u,} \label{pr-4}
\end{array}
\end{equation}
therefore
$$
{\bf P_0}{\bf P_{1,2}}(\lambda u)=\lambda {\bf P_0}{\bf P_{1,2}}u.
$$
\end{Remark}

\section{Matrix $P_2$ Reloaded}\label{Sec:matrices}

Now, let us explain how to lower the power of differential Equation (\ref{2.7}). We begin by noticing that the first integral of (\ref{2.3}) would be its Wro\'nskian, i.e.,
\begin{equation}
W_2(u,v)=C.
\label{3.1}
\end{equation}

After we remove function $v$ from (\ref{3.1}), we end up with a new {\bf third} order equation:
\begin{equation}
u'''=\frac{3u'u''}{u}-2xu'+(1-2C)u
\label{3.2}
\end{equation}
and it is straightforward to show that any solution of (\ref{3.2}) will serve as such for (\ref{2.7}) as well.

We are not finished. The 3rd order equation is homogeneous with respect to combination $\{u, u', u'', u'''\}$, which implies that the Cole--Hopf transformation $q=(\log u)'$ will further reduce it to a {\bf second} order O.D.E.:
\begin{equation}
q''=2q^3-2xq-2\alpha,
\label{3.3}
\end{equation}
where $\alpha=C-1/2$. It is this equation which we will subsequently call $P_2$, and for a good reason: it is reducible to the canonic form \eqref{P_2} via two simple substitutions:
\beqn
\displaystyle{x \to -2^{-1/3}x ,\qquad q \to -2^{1/3} w .}
\enqn

However, in order to prevent an unnecessary cluttering of our formulas by throwaway coefficients, for now we will continue working with (\ref{3.3}) under the assumption $q=q(x,\alpha)$.

It is both important and interesting to note that the ${\bf P_i}$ symmetries are in fact the famous B\"acklund transformations for $P_2$. Since we have chosen to embrace a non-canonical form of $P_2$, it is instructive to show these transformations. Here they are:
\begin{equation}
q(x,\alpha)\to {\bf P_0}q(x,\alpha)\equiv q_0(x,-\alpha)=-q(x,\alpha),
\label{3.4}
\end{equation}
\begin{equation}
q(x,\alpha)\to {\bf P_1}q(x,\alpha)\equiv q_1(x,\alpha+1)=-q(x,\alpha)+\frac{2\alpha+1}{q'(x,\alpha)+q^2(x,\alpha)-x},
\label{3.5}
\end{equation}
\begin{equation}
q(x,\alpha)\to {\bf P_2}q(x,\alpha)\equiv q_2(x,\alpha-1)=-q(x,\alpha)-\frac{2\alpha-1}{q'(x,\alpha)-q^2(x,\alpha)-x}.
\label{3.6}
\end{equation}

The transformations (\ref{3.4})--(\ref{3.6}) are well-known and well-studied, although they are usually derived from the B\"acklund transformations for KdV equations with the aid of a self-similar change of variables. We can see now that (\ref{3.4})--(\ref{3.6}) serve as discrete symmetries for $P_2$, i.e., the explicitly invertible B\"aclund transformations for (\ref{2.3}), which are, in turn, nothing but ST for the trivial NLS deformation.  Additionally, their connection with the Toda chain becomes very straightforward.

The (\ref{3.5}) and (\ref{3.6}) are well-studied in the literature, so we can now safely leave them and move on to the Jordan generalizations. As a particular example, consider the following matrix equation $P_2$:
\begin{equation}
W''=2W^3+xW+\alpha,
\label{3.7}
\end{equation}
where $W(x,\alpha)$ and $\alpha={\rm const}$ are $N\times N$ matrices. Our goal is to derive (\ref{3.7}) in the same manner as (\ref{3.3}) before that. To this end, we write a matrix equivalent of (\ref{2.3}):
\begin{equation}
U''=2UVU+xU,\qquad V''=2VUV+xV,
\label{3.89}
\end{equation}
where $U$ and $V$ are, again, $N\times N$ square matrices. The sought after analogue of ST Equations (\ref{2.5}) and (\ref{2.6}) (the analogous symmetries for the Davey--Stewartson equation have been studied in \cite{Tema1} by one of the authors) will be
\begin{equation}
\begin{split}
U\to U_1 &=V^{-1},\\
V\to V_1 &=V'V^{-1}V'-VUV-xV,
\label{3.10}
\end{split}
\end{equation}
\begin{equation}
\begin{split}
U\to U_{-1}&=U'U^{-1}U'-UVU-xU,\\
V\to V_{-1}&=U^{-1}.
\label{3.11}
\end{split}
\end{equation}

Note that the symmetries (\ref{3.10}) and  (\ref{3.11}) are mutually inverse:
$$
(U_1)_{-1}=(U_{-1})_1=U,\qquad (V_1)_{-1}=(V_{-1})_1=V.
$$

By removing $V$, we end up with a matrix analogue of (\ref{2.7}):
\begin{equation}
\begin{split}
U''''&-3U''U^{-1}U''+2U'U^{-1}U'U^{-1}U''+2U''U^{-1}U'U^{-1}U'-2U'U^{-1}U'''-\\
&-2U'''U^{-1}U'+2U'U^{-1}U''U^{-1}U'+2U'-2xU'U^{-1}U'+2xU''=0,
\label{2.77}
\end{split}
\end{equation}
whose order is reducible via the first integral
\begin{equation}
VU'-V'U=A={\rm const}.
\label{3.12}
\end{equation}

\begin{Remark} \label{Rem:AB} One can also pick an alternative integral: $U'V-UV'=B$. However, since the system \eqref{3.89} is invariant with respect to inversion $U \leftrightarrow V$, this option simply means that $B=-A$.
\end{Remark}

Now, we define the matrix-valued function $q$ as
\begin{equation}
q\equiv U'U^{-1}.
\label{3.13}
\end{equation}
such that it satisfies the condition
$$
V=(2U)^{-1}\left(q'+q^2-xE\right),
$$
so after a few simple calculations (\ref{3.12}), we obtain
\begin{equation}
q''=2q^3-2xq+E-2UAU^{-1}.
\label{3.14}
\end{equation}

Of course, if $A$ is proportional to a unitary matrix $A=cE$, then (\ref{3.14}) will be identical to \eqref{3.3}, except that $q$ will be an $N\times N$ matrix, whereas $2\alpha=2c-1$ will be equal to a product of $E$ and a number. The equation then turns into (\ref{3.7}) via the same point transformation as the one used in the scalar case.

It has been previously pointed out by Balandin in \cite{Balandin} that the diagonality of $A$ might be a necessary condition for the integrability of matrix $P_2$. We now see that the non-diagonality actually leads to the integrability (for our purposes, the {\em integrability} of a nonlinear equation means the existence of an explicitly invertible B\"acklund transformation---an analogue of the Schlesinger transformations) of some non-local integro-differential matrix analogue of $P_2$. However, this equation might be rewritten in a completely local form via a different substitution:
$$
Q\equiv U^{-1}U'.
$$
which results in a different, and rather elegant, matrix O.D.E.:
\begin{equation}
Q''=2Q^3-2xQ+[Q',Q]+E-2A.
\label{3.15}
\end{equation}

In a scalar case, the commutator vanishes, delivering us $P_2$.

\begin{Remark} \label{Rem:Sokolov}
This work was reaching its conclusion when we learned of a paper by Adler and Sokolov \cite{Adler-Sok}, published in 2021. There, the authors introduced three matrix-valued integrable generalizations of $P_2$, denoted by Adler and Sokolov as $P_2^0$, $P_2^1$, and $P_2^1$. Out of these three, the first one was (\ref{3.15}) exactly, whereas the second generalization can be derived from \eqref{3.15} via a specific matrix shift of an independent variable. Interestingly, the authors of \cite{Adler-Sok} have introduced their versions of equations mostly voluntarily (essentially acting on the principle of ``in a scalar case a term with the commutator vanishes; let's add it to the mix!''), while in our approach, \eqref{3.15} is a necessary outcome of previous calculations. Nevertheless, we admit that we cannot claim the honour of discovering \eqref{3.15}; it rightfully belongs to Adler and Sokolov.
\end{Remark}

To conclude our discussion of Jordan generalizations, let us jot down the B\"acklund transform for (\ref{3.15}). Let
$$
J\equiv Q^2+Q'-xE,
$$

Then, the direct B\"acklund transformation will have the form
\begin{equation}
Q\to Q_1=JQJ^{-1}-J'J^{-1},\qquad A\to A_1=A+E,
\label{MB1}
\end{equation}
while its inverse
\begin{equation}
Q\to Q_{-1}1=I^{-1}QI+I^{-1}I',\qquad A\to A_{-1}=A-E,
\label{MB2}
\end{equation}
where
$$
I=Q^2-Q'-xE=J-2Q'.
$$

\section{The Vectorial Painlev\'e Equation}\label{Sec:vectors}

Now, let us take one more step and attempt to tackle a {\em vectorial} generalization of $P_2$. In order to do this, we will use the model of vector NLS originally introduced by Sokolov (and studied in \cite{Svin}).
Let ${\bf U}=\{u_1(x),u_2(x),..,u_{_N}(x)\}^T$ and ${\bf V}=\{v_1(x),v_2(x),..,v_{_N}(x)\}^T$ be two $N$-dimensional vectors with a standard Euclidean scalar product
$$
({\bf U}{\bf V})=\sum_{i=1}^N u_iv_i.
$$

What kind of equation are we going to call an $N$-component vector $P_2$? It will be a system of $2N$ equations, with the following vector representation:
\begin{equation}
\begin{split}
{\bf U}''&=4({\bf U}{\bf V}){\bf U}-2({\bf U}{\bf U}){\bf V}+x{\bf U},\\
{\bf V}''&=4({\bf U}{\bf V}){\bf V}-2({\bf V}{\bf V}){\bf U}+x{\bf V}.
\label{vec-1}
\end{split}
\end{equation}

One might at first hesitate calling (\ref{vec-1}) the $P_2$ equation, but it is in fact a literal generalization of $P_2$. For example, in the case of $N=2$, (\ref{vec-1}) has the form
\begin{equation}
\begin{split}
u_1''&=4(u_1v_1+u_2v_2)u_1-2(u_1^2+u_2^2)v_1+xu_1,\\
u_2''&=4(u_1v_1+u_2v_2)u_2-2(u_1^2+u_2^2)v_2+xu_2,\\
v_1''&=4(u_1v_1+u_2v_2)v_1-2(v_1^2+v_2^2)u_1+xv_1,\\
v_2''&=4(u_1v_1+u_2v_2)v_2-2(v_1^2+v_2^2)u_2+xv_2,
\label{vec-2}
\end{split}
\end{equation}
and for $N=3$,
\begin{equation}
\begin{split}
u_1''&=4(u_1v_1+u_2v_2+u_3v_3)u_1-2(u_1^2+u_2^2+u_3^2)v_1+xu_1,\\
u_2''&=4(u_1v_1+u_2v_2+u_3v_3)u_2-2(u_1^2+u_2^2+u_3^2)v_2+xu_2,\\
u_3''&=4(u_1v_1+u_2v_2+u_3v_3)u_3-2(u_1^2+u_2^2+u_3^2)v_3+xu_3,\\
v_1''&=4(u_1v_1+u_2v_2+u_3v_3)v_1-2(v_1^2+v_2^2+v_3^2)u_1+xv_1,\\
v_2''&=4(u_1v_1+u_2v_2+u_3v_3)v_2-2(v_1^2+v_2^2+v_3^2)u_2+xv_2,\\
v_3''&=4(u_1v_1+u_2v_2+u_3v_3)v_3-2(v_1^2+v_2^2+v_3^2)u_3+xv_3.
\label{vec-3}
\end{split}
\end{equation}

Consider (\ref{vec-2}). Solve the first two equations for $v_1$, $v_2$ and substitute them into the remaining equation. The result will be a system of two equations of fourth order for $u_1$ and $u_2$ that we will omit here owing to its cumbersomeness. However, if we assume that $u_1=u_2=u$, both of these equations will be reducible to  (\ref{2.7}), which is just another form of $P_2$. In a similar vein, the system (\ref{vec-3}) can be coaxed into producing a system of three equations of fourth order with a total of 92 terms in each; they are tamed by the conditions $u_1=u_2=u_3=u$, contracting down to a single Equation (\ref{2.7}).

The vector form of $P_2$ permits an explicit self-transformation of the kind
\begin{equation}
\begin{split}
{\bf U}\to {\bf U}^{(1)}&=-\frac{({\bf U}{\bf U})'}{({\bf U}{\bf U})}{\bf U}'+\frac{({\bf U}{\bf U})''}{2({\bf U}{\bf U})}{\bf U}-({\bf U}{\bf U}){\bf V},\\
{\bf V}\to{\bf V}^{(1)}&=-\frac{{\bf U}}{({\bf U}{\bf U})},
\label{vecB1}
\end{split}
\end{equation}
\begin{equation}
\begin{split}
{\bf U}\to{\bf U}^{(-1)}&=-\frac{{\bf V}}{({\bf V}{\bf V})},\\
{\bf V}\to {\bf V}^{(-1)}&=-\frac{({\bf V}{\bf V})'}{({\bf V}{\bf V})}{\bf V}'+\frac{({\bf V}{\bf V})''}{2({\bf V}{\bf V})}{\bf V}-({\bf V}{\bf V}){\bf U}.
\label{vecB2}
\end{split}
\end{equation}
which allows us to construct infinitely diverse families of exact solutions for (\ref{vec-1}) based on the initial trivial offering. For example, if we take an arbitrary constant $N$-vector
$$
{\bf w}=\{w_1,w_2,w_3,...,w_{_N}\}^T,
$$
then the simplest solutions of (\ref{vec-1}) will be of the form
\begin{equation}
{\bf V}={\bf w},\qquad {\bf U}=-\frac{{\bf w} x}{2|{\bf w}|^2},
\label{sol}
\end{equation}

Applying to them our transformation (\ref{vecB1}), after the $n$-th iteration, we will gather the following solution:
\begin{equation}
{\bf V}^{(n)}={\bf w}v^{(n)},\qquad {\bf U}^{(n)}=u^{(n)}\frac{{\bf w}}{|{\bf w}|^2},
\label{los}
\end{equation}
where the scalar functions $v^{(n)}=v^{(n)}(x)$, $u^{(n)}=u^{(n)}(x)$ are derived with the aid of the Schlesinger transformation:
\begin{equation}
\begin{split}
u^{(n+1)}&=u^{(n)}\left(\log u^{(n)}\right)''-\left(u^{(n)}\right)^2v^{(n)},\\
v^{(n+1)}&=-\frac{1}{u^{(n)}}.
\label{PrSl}
\end{split}
\end{equation}

Here are the fruits of our labours. Applying (\ref{PrSl}) for (\ref{sol}) yields
\begingroup\makeatletter\def\f@size{8.5}\check@mathfonts
\def\maketag@@@#1{\hbox{\m@th\normalsize \normalfont#1}}%
\begin{equation}
\begin{split}
u^{(1)}&=\frac{2-x^3}{4x},\qquad  \qquad \qquad v^{(1)}=\frac{2}{x},\\
u^{(2)}&=\frac{x^6-10x^3-20}{8(2-x^3)},\qquad \qquad v^{(2)}=\frac{2x}{x^3-2},\\
u^{(3)}&=-\frac{x(x^9-30x^6-1400)}{16(x^6-10x^3-20)},\qquad v^{(3)}=\frac{8(x^3-2)}{x^6-10x^3-20},\\
u^{(4)}&=-\frac{x^{15}-70x^{12}+700x^9-9800x^6-196000x^3+196000}{32x(x^9-30x^6-1400)},\\
 v^{(4)}&=\frac{16(x^6-10^3-20)}{x(x^9-30x^6-1400) },\\
u^{(5)}&=-\frac{x^{21}-140x^{18}+4620x^{15}-78400x^{12}-1078000x^9-45276000x^6+301840000(x^3+1)}{64(x^{15}-70x^{12}+700x^9-9800x^6-196000x^3+196000)} ,\\
 v^{(5)}&=\frac{32x(x^9-30x^6-1400)}{x^{15}-70x^{12}+700x^9-9800x^6-196000x^3+196000},
\label{sol5}
\end{split}
\end{equation}
\endgroup
etc. Thus, we obtain a chain of exact solutions $u^{(k)}$, $k \in \N$, which we shall recognize straight away, because we have seen them before. In fact, they are almost identical to the $u_k$ solutions \eqref{u1-u6} we procured in Section \ref{Sec:deformations}. A closer examination reveals the following relationship between the old solutions $u_k$ and the new one:
\beqn
v^{(k)} = (-1)^k \cdot u_{_{k}}^{^{(-1)^k}}, \qquad u^{(k)} = (-1)^k \cdot u_{_{k+1}}^{^{(-1)^k}}.
\enqn

Another interesting starting point for generating new solutions would be the null-vector $V\equiv 0$. This choice not only converts the system \eqref{vec-1} into a linear equation, but it also implies that every component $u_k$ of vector $U$ satisfies the {\em Airy equation}:
\beq \label{Airy_eq}
u_k'' = x ~u_k.
\enq

Two linearly independent solutions of \eqref{Airy_eq} are the so-called {\em Airy functions} $\Ai(x)$ and $\Bi(x)$, which for $x\in\R$ can be written as \cite{Abramowitz}
\beq \label{Ai-Bi}
\begin{split}
\Ai(x) &= \frac{1}{\sqrt{\pi}} \int\limits_{0}^\infty \cos\left(\frac{t^3}{3} + x t\right) dt \\
\Bi(x) &= \frac{1}{\sqrt{\pi}} \int\limits_{0}^\infty \left[\exp\left(-\frac{t^3}{3} + x t\right) + \sin\left(\frac{t^3}{3} + x t\right) \right]dt.
\end{split}
\enq

One can therefore write general solutions of Equation \eqref{Airy_eq} as a linear combination of $\Ai(x)$ and $\Bi(x)$:
\beq \label{uAi}
u_k = a_k \Ai(x) + b_k \Bi(x), \qquad a_k, b_k \in \R,
\enq
and applying the transformations \eqref{vecB1} and \eqref{vecB2} to \eqref{uAi} (keeping in mind that $v_k \equiv 0$), we will end up with a rather different family of exact solutions that not only contains $2N$ arbitrary parameters $\{a_k\}_{k=1}^N$, $\{b_k\}_{k=1}^N$, but is also completely entirely determined via the Airy functions and their derivatives. (This is but one of many examples where the solutions of the integrable nonlinear differential equations can be written in terms of the Airy functions; for instance, in \cite{Yurova}, the Airy functions are shown to be an important ingredient in the solution of the Cauchy problem for the Novikov--Veselov equation.)

Before the conclusion, let us emphasize the fact that  (\ref{vec-1}) is a JP-system, whose Jordan algebra $V_{_N}$ is determined by the Jordan product
$$
({\bf e},{\bf x}){\bf y}+({\bf e},{\bf y}){\bf x}-({\bf x},{\bf y}){\bf e},
$$
where ${\bf e}$, ${\bf x}$, and ${\bf y}$ are the elements of $N$-dimensional vector space; $(\,.\,,\,.\,)$ is the scalar product; and ${\bf e} = (1,0,0,..)^T$ is a unit in the $V_{_N}$ algebra.

\section{On a Way to the Painlev\'e IV} \label{Sec:P4}

In the previous sections, we have succeeded in developing a technique for the construction of solutions for the $P_2$ equation, thus completing the main goal of this article. Nevertheless, the success motivates us to at least take a short glance at the possibility of extending our technique for the remaining five Panilev\'e equations. Thus, in this section we will briefly explain how to do that for Painlev\'e IV. First of all, we have to point our that NLS, whose deformations acted as a sort of a launchpad for our investigation, is tightly related to yet another famous integrable equation known as the Kadomtsev--Petviashvili equation (KP). This relationship, discovered and used in \cite{Mat1,Mat2} to generate new rogue wave soliton solutions (i.e., the solutions that are localized in both space and time; originally discovered by Howell Peregrine in 1983 for NLS \cite{Pereg}, the rogue wave solutions have been steadily cropping up in almost every field of mathematical physics, from the collapse of intrathermocline eddies in the ocean \cite{Yurova14} to a magnetic ``impacton'' arising during a collision of two positon solutions in ferromagnetic nanowires \cite{YY-NLS}), can only be properly understood in the framework of the AKNS hierarchy (see Remark \ref{Rem:AKNS} and article \cite{MS}). For our purposes, it is important to single out the existence of an explicit relationship between the solutions of (split) NLS and KP. The latter equation can be written as a compatibility condition of two different linear equations; their shapes determine two types of dressing chains of discrete symmetries. Say we choose a first one \cite{YurovTMF}:
\begin{equation}
\alpha(s-f)_y+(s+f)_{xx}+s_x^2-f_x^2=0,
\label{y-Ch}
\end{equation}
with $\alpha^2=\pm 1$. Let us add a periodic boundary condition
$$
f_{n+N}=f_n+c(y),
$$
where $f=f_n$, $s=f_{n+1}$. For a period $N=3$, choose $c=-2y/\alpha$ and introduce new field variables $g_n$,
$n=1,2,3$ as follows:
$$
f_1=\frac{1}{2}\left(g_1- g_2+g_3+c\right),\qquad
f_2=\frac{1}{2}\left(g_1 + g_2-g_3-c\right),\qquad
f_3=\frac{1}{2}\left(-g_1+g_2+g_3+c\right).
$$

This will result in three equations for our new functions $g_n$; their forms have been previously explicitly derived in \cite{Tema2}. Excluding $g_3$ and using the	compatibility condition
\mbox{$\partial_y \partial^2_x ~g_{2}=\partial_x^2 \partial_y ~g_{2}$} will produce for us the nonlinear equation, which, after some simple transformations, can be written as
\begin{equation}
\begin{array}{l}
\displaystyle{
z_{xx}=\frac{1}{2}\frac{z_x^2}{z}+\frac{3}{2}z^3+4xz^2+2(x^2-2)z+
\frac{\beta}{z}+}\\
\\
\displaystyle{
+\frac{3\alpha^2q^2}{{2z}}-3\alpha qz+
\frac{\alpha}{z}D_x^{-1}\left(z^3+2xz^2-3\alpha qz\right)_y ~,\qquad
z_y=q_x.}
\end{array}
\label{188}
\end{equation}
where $z={{\it g_{1}}_{{x}}}$ and $q={{\it g_{1}}_{{y}}}$. In a one-dimensional limit, the dependence on $y$ vanishes along with all the terms containing $q$ and the derivatives with respect to $y$, so Equation (\ref{188}) ends up being the sought after $P_4$.

If we repeat all these calculations, this time starting out from the Jordan generalization of NLS, we shall arrive at the Jordan generalizations of $P_4$, i.e., at the JP-systems. We will return to this problem in a subsequent paper.

 \section{Conclusions}\label{Sec:conclusion}

In this article, we have demonstrated how the coupling of NLS deformations with the Schlesinger transformation naturally produces the Second Painlev\'e equation $P_2$ and its B\"acklund transformations. The resulting scheme is versatile enough to help generalize the JS-system formalism for the {\em JP-systems}, by which we understand the multiple-fields integrable generalization of $P_2$, based upon the unital Jordan algebras. This opens up a number of very promising  avenues of research, including the ultimate goal of classification  and description of JP-systems. Another interesting possibility lies in applying the new results to the task of constructing new discrete analogues of $P_2$ (cf. \cite{Diskr}), associated with the Jordan algebras. The novel approach developed herein is based on the invertible B\"acklund transformations, which serve as a crucial part of the JP-systems theory. This further reinforces our opinion that the process of the discretization of the JP-system will see no substantial difficulties.

Now let us summarize the main results.
\begin{enumerate}

\item We have demonstrated how the trivial deformations of NLS, invariant with respect to the Schlesinger transformations, can be reduced to a system of two split O.D.E.s of second order that inherits this symmetry.

\item The resulting system is shown to be equivalent to a single fourth order O.D.E.; its order can be reduced, producing a familiar equation: $P_2$.

\item Inherent Schlesinger symmetries appear to be nothing more but well-known explicitly invertible B\"acklund transformations for $P_2$. Thus, we obtain a very simple and concise method for the derivation of both $P_2$ and its B\"acklund autotransformations.

\item The simplicity of the method paves the way for its generalization for multiple fields models---a problem which is traditionally considered a difficult one. We demonstrate how the new approach dispels the difficulties in establishing multicomponent $P_2$ generalizations associated with arbitrary unital Jordan algebras. This implies that the multicomponent integrable (in the sense of having an analogue of the Schlesinger transformation) $P_2$ generalizations can actually be classified, for example, by a set of structural constants of a corresponding Jordan algebra. We have called such models {\em JP-systems}.

\item Another interesting observation that needs to be emphasized is that existence of explicitly invertible B\"acklund transformations opens a way to generalize the proposed method to the {\em discrete} Painlev\'e equations. Taking into account the rapid growth of interest in the discretization of Painlev\'e equations (see \cite{Diskr,Dis-2,Dis-3,Miron}), it would not be unreasonable to expect the said method to be useful to the research.
\end{enumerate}

One last question must be addressed before we wrap up. In this article, we have developed a method of generalization for the {\em Second} Painlev\'e equation. What about the remaining five? In the last Section, we briefly touched upon Painlev\'e IV and how it can be naturally introduced into our considerations via the fascinating relationship between NLS and KP equations. As for the rest, at this stage we can only surmise that it might be possible to construct a similar scheme for at last some of them, since these equations also posses the Schlesinger symmetries and the chains of discrete symmetries \cite{Tema2}. This, however, is of course a matter for another time and another article. For now, let us simply conclude by stating out humble hope that the approach developed in this article might be of use in both the theory of Painlev\'e equations and in mathematical physics in general.

\section*{Acknowledgments}
The article was supported by the Ministry of Science and Higher Education of the Russian Federation (agreement no. 075-02-2021-1748).


\end{document}